\documentclass[fleqn,twoside,11pt]{article}
\usepackage{latexsym} 

\newcommand{\be}{\begin{equation}}
\newcommand{\ee}{\end{equation}}
\newcommand{\ba}{\hspace*{-5pt}\begin{array}}
\newcommand{\ea}{\end{array}}

\newcommand{\rank}{\mathop{\rm rank}\nolimits}
\newcommand{\const}{\mathop{\rm const}\nolimits}
\newcommand{\deter}{\mathop{\rm det}\nolimits}

\begin{document}

\begin{flushleft}
\Large \bf On Reduction and {\bfseries \itshape Q}-conditional
(Nonclassical) Symmetry
\end{flushleft}

\noindent
{\bf Roman O. POPOVYCH}

\bigskip

\noindent
{\small\it Institute of  Mathematics of
the  National  Ukrainian Academy of Sciences, \\
3 Tereshchenkivska Str. 3, 01601 Kyiv,  Ukraine}

\bigskip

\noindent  {\small URL:
{\tt http://www.imath.kiev.ua/\~{}rop/} \\
E-mail: {\tt rop@imath.kiev.ua}}

\begin{abstract}
\noindent
Some aspects of $Q$-conditional symmetry and of its connections 
with reduction and compatibility are discussed.
\end{abstract}

\vspace{1ex}


\bigskip

There are a number of examples in the history of science when a scientif\/ic 
theory quickly developed, gave good results and applications, and 
had no suf\/f\/iciently good theoretical fundament. Let us remember 
mathematical analysis in the XVIII-th century. It is true in some
respect for the theory of $Q$-conditional (also called non-classical) 
symmetry. The pioneer paper of Bluman and Cole \cite{popov_r:bluman&cole} 
where this conception appeared was published 28 years ago. And even now,
some properties of $Q$-conditional symmetry are not completely investigated.

\medskip

\noindent
{\bf 1.} $Q$-conditional symmetry being a generalization of the classical
Lie symmetry, we recall brief\/ly some results of the Lie theory.

Let 
\be \label{popov_r:general.system}
L(x, {\mathop {u}\limits_{(r)}})=0, \quad L=(L^1, L^2, \ldots, L^q),
\ee
be a system of $q$ partial dif\/ferential equations (PDEs) of order $r$ 
in $n$ independent variables $x=(x_1,x_2,\ldots,x_n)$
and $m$ dependent variables $u=(u^1,u^2,\ldots,u^m),$ 
satisfying the maximal rank condition. 
Here $\mathop u\limits_{(r)}$ denotes all  derivatives of 
the function $u$ up to order~$r$.

Consider a dif\/ferential operator of the f\/irst order
\[
Q=\sum_{i=1}^n\xi^i(x,u)\partial_{x_i}
 +\sum_{j=1}^m\eta^j(x,u)\partial_{u^j}.
\]
Actions of the operator $Q$ 
on the functions $u^j$, $j=\overline{1,m},$
are def\/ined by the formula
\[
Qu^j=\eta^j-\sum_{i=1}^n\xi^iu^j_i, \quad j=\overline{1,m}.
\]

\noindent
{\bf Def\/inition.} {\it An operator $Q$ is called a Lie symmetry operator 
of  system (\ref{popov_r:general.system}) if  system (\ref{popov_r:general.system}) 
is invariant under the local transformations generated by the
operator $Q.$} 
\medskip

\noindent
{\bf Theorem.}
{\it An operator $Q$ is a Lie symmetry operator 
of  system (\ref{popov_r:general.system}) if and only if
the condition
\be \label{popov_r:invariance.condition}
\left.\ba{c}\mathop Q\limits_{(r)}L(x, \mathop
u\limits_{(r)})\\ \ea\!
\right|_{\:\overline{\{L(x, \mathop u\limits_{(r)})=0\}}\:=:\:K}
=0 
\ee
is satisfied where $\mathop Q\limits_{(r)}$ denotes the $r$-th prolongation 
of the operator $Q$ and $K$ does the set of  differential consequences 
of  system (\ref{popov_r:general.system}). }

\medskip

\noindent
{\bf Theorem.}
{\it The set of  Lie symmetry operators of  system
(\ref{popov_r:general.system}) 
is a Lie algebra under the standard Lie brackets of differential operators 
of the first order. }

\medskip

Consider the set of the Lie symmetry operators 
$Q_1,Q_2,\ldots,Q_s$ $(s<n)$
of  system (\ref{popov_r:general.system})
which satisfy the following conditions:
\begin{enumerate}
\item
$\langle Q_1,\: Q_2,\dots,\: Q_s\rangle $ is a Lie algebra; 
\item
$\rank||\xi^{ki}||\,{}_{k=1}^s\,{}_{i=1}^n=
\rank||\xi^{ki},\eta^{kj}||\,{}_{k=1}^s\,{}_{i=1}^n\,{}_{j=1}^m=s.
$
\end{enumerate}
Then, a general solution of the system 
\be \label{popov_r:ansatz.eq}
Q^ku^j=0, \quad  k=\overline{1,s}, \quad j=\overline{1,m},
\ee
can be written in the form 
\be \label{popov_r:proansatz}
W^j(x,u)=\varphi^j(\omega(x,u)), \quad 
j=\overline{1,m}, \quad \omega=(\omega_1,\omega_2,\ldots,\omega_{n-s}),
\ee
where $\varphi^j,$ $j=\overline{1,m},$ 
are arbitrary dif\/ferentiable functions of $\omega;$ 
$W^j,$ $j=\overline{1,m},$ and
$\omega^l,$ $l=\overline{1,n-s},$ are functionally 
independent f\/irst integrals of  system (\ref{popov_r:ansatz.eq}) 
and the Jacobian of $W$ under $u$ does not vanish, that is, 
\be \label{popov_r:condition.existance.ansatz}
\deter ||W^j_{u^{j'}}||\,_{j,j'=1}^m\not=0.
\ee
Owing to  condition (\ref{popov_r:condition.existance.ansatz}) being satisf\/ied,
we can solve  equation (\ref{popov_r:proansatz}) with respect to $u.$ 
As a result, we obtain a form to f\/ind $u$, also called "ansatz":
\be \label{popov_r:general.ansatz}
u=F(x,\varphi(\omega)),
\ee
where $u,$ $F,$ and $\varphi$ have $m$ components.

\medskip

\noindent
{\bf Theorem.}
{\it Substituting (\ref{popov_r:general.ansatz}) into (\ref{popov_r:general.system}), 
we obtain a set of equations that is equivalent to a system for the
functions $\varphi^j$ ($j=\overline{1,m}$), 
depending only on variables $\omega_1,$ $\omega_2,$ \dots, $\omega_{n-s}$ 
(that is called the reduced system corresponding to 
the initial system (\ref{popov_r:general.system}) and the algebra 
$\langle Q_1,\: Q_2,\dots,\: Q_s\rangle $). }

\medskip

Let us emphasize some properties of Lie symmetry and Lie reduction.
\begin{enumerate}
\item
The {\em local approach,} that is, a system of PDEs is assumed 
a manifold in a prolongated space.  
\item
To f\/ind the maximal Lie invariance algebra of a system of PDEs, 
we have to consider {\em all the non-trivial differential consequences}  
of this system of order less than $r+1.$
\item
{\em Conditional compatibility,} that is, the system 
\[
L(x, \mathop u\limits_{(r)})=0, \quad
Q^ku^j=0, \quad  k=\overline{1,s}, \quad j=\overline{1,m},
\]
consisting of an inverstigated system od PDEs and the surface invariant 
conditions is compatible if and only if the corresponding reduced system 
is compatible.
\end{enumerate}
Note that there are a number of examples when the reduced system is 
not compatible. For instance, consider the equation $tu_t+xu_x=1.$ 
It is invariant under the dilatation operator 
$t\partial_t+x\partial_x,$ but the corresponding reduced system ($0=1$) 
is not compatible.

\medskip

\noindent
{\bf 2.} A direct generalization of Lie symmetry is conditional symmetry 
introduced by Prof. Fushchych in 1983 \cite{popov_r:fushchych.1983}.

\medskip

\noindent
{\bf Def\/inition.}
{\it Consider a system of the form (\ref{popov_r:general.system}) and 
an appended additional condition
\be \label{popov_r:additional.condition}
L'(x, \mathop u\limits_{(r')})=0.
\ee
System (\ref{popov_r:general.system}), (\ref{popov_r:additional.condition}) being
invariant under an operator $Q,$
 system (\ref{popov_r:general.system}) is called conditionally
invariant under this operator. }

\medskip

Conditional symmetry was a favourite conception of Prof.~Fushchych.

\medskip

\noindent
{\bf 3.} Let us pass to $Q$-conditional symmetry. Below we take $m=q=1,$ i.e.,
we consider one PDE in one unknown function. 

It can be seemed on the face of it that $Q$-conditional (non-classical) 
symmetry is a particular case of conditional symmetry. 
One often uses the following def\/inition of non-classical symmetry.

\medskip

\noindent
{\bf Def\/inition 1.}
{\it A PDE of the form (\ref{popov_r:general.system}) is called $Q$-conditionally 
(or non-classical) invariant under an operator $Q$
if the system of  equation (\ref{popov_r:general.system}) and the 
invariant surface condition
\be \label{popov_r:invariant.surface.condition}
Qu=0 
\ee
is invariant under this operator. }

\medskip

It is not a quitely correct def\/inition. Indeed, 
equation (\ref{popov_r:general.system}) is $Q$-conditionally invariant 
under an operator $Q$ in terms of def\/inition 1 if the following 
condition is satisf\/ied:
\be \label{popov_r:Q.invariance.condition.1}
\left.\ba{c}\mathop Q\limits_{(r)}L(x, \mathop u\limits_{(r)})\\ \ea\!
\right|_{\:\overline{\{L(x, \mathop u\limits_{(r)})=0,\; Qu=0\}}\:=:\:M}
=0, 
\ee
where $M$ denotes the set of  dif\/ferential consequences 
of  system (\ref{popov_r:general.system}), (\ref{popov_r:invariant.surface.condition}).
But the equation $\mathop Q\limits_{(r)}L(x, \mathop u\limits_{(r)})$
belongs to $M$ because it is simple to see that the following formula 
is true:
\[
\mathop Q\limits_{(r)}L(x, \mathop u\limits_{(r)})=
\sum_{\alpha:\:|\alpha|\le r}
\left(\partial_{u_\alpha}L(x, \mathop u\limits_{(r)})\right)
\cdot D_\alpha(Qu)+
\sum_{i=1}^n\xi^iD_{x_i}\left(L(x, \mathop u\limits_{(r)})\right),
\]
where $\alpha=(\alpha_1,\alpha_2,\ldots,\alpha_n)$ is a multiindex, 
$D$ is the full derivation operator. Therefore, equation 
(\ref{popov_r:Q.invariance.condition.1}) is an identity for an arbitrary 
operator $Q.$

We like the slightly dif\/ferent def\/inition of $Q$-conditional symmetry that 
appeared in \cite{popov_r:fushchych&tsyfra} and was developed in 
\cite{popov_r:zhdanov&tsyfra}.

Following \cite{popov_r:zhdanov&tsyfra},
 at f\/irst, let us formulate two auxiliary def\/initions.

\medskip

\noindent
{\bf Def\/inition.}
{\it Operators $Q^k,$ $k=\overline{1,s}$ form an involutive set 
if there exist functions $f^{klp}=f^{klp}(x,u)$ satisfying the condition
\[
[Q^k,Q^l]=\sum_{p=1}^sf^{klp}Q^p, \quad {k=\overline{1,s}}, 
\quad \mbox{where}\quad [Q^k,Q^l]=Q^kQ^l-Q^lQ^k.
\]}

\noindent
{\bf Def\/inition.}
{\it Involutive sets of operators $\{Q^k\}$ and $\{\tilde Q^k\}$  
are equivalent if
\[
\exists\,\lambda_{kl}=\lambda_{kl}(x,u)\;
\bigl(\deter||\lambda_{kl}||\,_{k,l=1}^s\not=0 \bigr):\quad
\tilde Q^k=\sum_{l=1}^s\lambda_{kl}Q^l, 
\quad {k=\overline{1,s}}. 
\]}

\noindent
{\bf Def\/inition 2 \cite{popov_r:zhdanov&tsyfra}.}
{\it The PDE (\ref{popov_r:general.system}) is called $Q$-conditionally 
invariant under the involutive sets of operators $\{Q^k\}$ 
if 
\be \label{popov_r:Q.invariance.condition.2}
\left.\ba{c}{\mathop Q\limits_{(r)}}^kL(x, \mathop u\limits_{(r)})\\ \ea\!
\right|_{\:L(x, \mathop u\limits_{(r)})=0,
\;\:\overline{\{Q^lu=0, \quad l=\overline{1,s}\}}\:=:\:N}
=0, \quad {k=\overline{1,s}},
\ee
where $N$ denotes the set of  differential consequences 
of the invariant surface conditions 
$Q^lu=0, \quad l=\overline{1,s},$ of order 
either less than or equal to $r-1.$}

\medskip

In fact, namely this def\/inition is always used to f\/ind $Q$-conditional 
symmetry. Restriction on the maximal order of the dif\/ferential consequences 
of the invariant surface conditions is not essential.

\medskip

\noindent
{\bf Lemma.}
{\it If  equation (\ref{popov_r:general.system}) is $Q$-conditionally
invariant under an involutive set of operators $\{Q^k\}$
then  equation (\ref{popov_r:general.system}) is $Q$-conditionally
invariant under an arbitrary involutive set of operators 
$\{\tilde Q^k\}$ being equivalent to the set $\{Q^k\}$ too.}

\medskip

It follows from Def\/inition 2 that $Q$-conditional symmetry conserves
the properties of Lie symmetry which  are emphasized above, and the 
reduction theorem is proved in the same way as for Lie symmetry. 

The conditional compatibility property is conserved too. Therefore, 
reduction under  $Q$-conditional symmetry operators does not 
always follow compatibility of the system from the initial equation 
and the invariant surface conditions. 
And vice versa, compatibility of this system does not follow 
$Q$-conditional invariance of the investigated equation under the 
operators $Q^k.$ For instance, the equation 
\be \label{popov_r:example1}
u_t+u_{xx}-u+t(u_x-u)=0
\ee
is not invariant under the translation operator with respect to $t$
$(\partial_t),$ and  system of the equation (\ref{popov_r:example1})
and $u_t=0$ is compatible because it has the non-trivial solution 
$u=Ce^x$ $(C=\const).$

Unfortunately, in terms of the local approach,  equation 
(\ref{popov_r:Q.invariance.condition.2}) is not a necessary condition of reduction 
with respect to the corresponding ansatz, although the conceptions 
of $Q$-conditional symmetry  and reduction are very similar. For example, 
the equation 
\[
u_t+(u_x+tu_{xx})(u_{xx}+1)=0
\]
is reduced to the equation $\varphi''+1=0$ by means of the ansatz 
$u=\varphi(x)$ and is not invariant under the operator $\partial_t.$
It is possible that  the def\/inition of $Q$-conditional invariance 
can be given in another way 
for the condition analogous to (\ref{popov_r:Q.invariance.condition.2})  
to be a necessary condition of reduction.

\medskip

\noindent
{\bf 4.} Last time, $Q$-conditional symmetry of a number of PDEs was
investigated, 
in particular, by Prof.~Fushchych and his collaborators 
(see \cite{popov_r:fshs} for references). 
Consider two simple examples.

At f\/irst, consider the one-dimensional linear heat equation
\be \label{popov_r:olhe} 
u_t=u_{xx}. 
\ee
Lie symmetries of  equation (\ref{popov_r:olhe}) are well known. 
Its maximal Lie invariance algebra is generated by the operators
\be \label{popov_r:mia.olhe}
\ba{l}
\partial_t, \quad
\partial_x, \quad
G=t\partial_x-\frac{1}{2}xu\partial_u, \quad
I=u\partial_u, \quad
D=2t\partial_t+x\partial_x, 
\\[1.5ex]
\Pi=4t^2\partial_t+4tx\partial_x-(x^2+2t)u\partial_u, \quad
f(t,x)\partial_u, 
\ea
\ee
where $f=f(t,x)$ is an arbitrary solution of (\ref{popov_r:olhe}). 
Firstly, $Q$-conditional symmetry of (\ref{popov_r:olhe}) was inverstigated 
by Bluman and Cole in \cite{popov_r:bluman&cole}.

\medskip

\noindent
{\bf Theorem 1 \cite{popov_r:fshsp}.}
{\it An arbitrary $Q$-conditional symmetry operator 
of the heat equation (\ref{popov_r:olhe}) is equivalent to either the operator
\[
Q=\partial_t+g^1(t,x)\partial_x+\bigl(g^2(t,x)u+g^3(t,x)\bigr)\partial_u,
\quad \mbox{where} 
\]
\be \label{popov_r:defining.equation.1}
g^1_t-g^1_{xx}+2g^1_xg^1+2g^2_x=0, \quad
g^k_t-g^k_{xx}+2g^1_xg^k=0,  \quad k=2,3,
\ee
or the operator
\[
Q=\partial_x+\theta(t,x,u)\partial_u,
\quad \mbox{where} 
\]
\be \label{popov_r:defining.equation.2}
\theta_t+\theta_{xx}+2\theta\theta_{xu}-\theta^2\theta_{uu}=0.
\ee}

The system of def\/ining equations (\ref{popov_r:defining.equation.1}) 
was f\/irstly obtained by Bluman and Cole \cite{popov_r:bluman&cole}. 
Further investigation of  system (\ref{popov_r:defining.equation.1}) 
was continued in \cite{popov_r:webb} where the question of linearization of 
the f\/irst two equations of (\ref{popov_r:defining.equation.1}) was
studied. The general solution of the problem of linearization 
of (\ref{popov_r:defining.equation.1}) and (\ref{popov_r:defining.equation.2}) 
was given in \cite{popov_r:fshsp}. We investigated Lie symmetry properties 
of (\ref{popov_r:defining.equation.1}) and (\ref{popov_r:defining.equation.2}).

\medskip

\noindent
{\bf Theorem 2 \cite{popov_r:fshsp}.}
{\it The maximal Lie invariance algebra (\ref{popov_r:defining.equation.1}) 
is generated by the operators
\be \label{popov_r:mia.defining.equation.1}
\ba{l}
\partial_t, \quad
\partial_x, \quad
G^1=t\partial_x+\partial_{g^1}-\frac{1}{2}g^1\partial_{g^2}
-\frac{1}{2}xg^3\partial_{g^3}, \quad
I^1=g^3\partial_{g^3}, \quad
\\[1.5ex]
D^1=2t\partial_t+x\partial_x-g^1\partial_{g^1}-2g^2\partial_{g^2}, 
\quad
(f_t+f_xg^1-fg^2)\partial_{g^3}, 
\\[1.5ex]
\Pi^1=4t^2\partial_t+4tx\partial_x-4(x-tg^1)\partial_{g^1}
-(8tg^2-2xg^1-2)\partial_{g^2}-(10t+x^2)g^3\partial_{g^3}, 
\ea
\ee
where $f=f(t,x)$ is an arbitrary solution of (\ref{popov_r:olhe}). }

\medskip

\noindent
{\bf Theorem 3 \cite{popov_r:fshsp}.}
{\it The maximal Lie invariance algebra of  (\ref{popov_r:defining.equation.2}) 
is generated by the operators
\be \label{popov_r:mia.defining.equation.2}
\ba{l}
\partial_t, \quad
\partial_x, \quad
G^2=t\partial_x+-\frac{1}{2}xu\partial_u
-\frac{1}{2}(x\theta+u)\partial_\theta, \quad
I^2=u\partial_u+\theta\partial_\theta, 
\\[1.5ex]
D^2=2t\partial_t+x\partial_x+u\partial_u, 
\quad
f\partial_u+f_x\partial_\theta, 
\\[1.5ex]
\Pi^2=4t^2\partial_t+4tx\partial_x-(x^2+2t)u\partial_u
-(x\theta+6t\theta-2xu)\partial_\theta, 
\\[1.5ex]
\ea
\ee
where $f=f(t,x)$ is an arbitrary solution of (\ref{popov_r:olhe}). }

\medskip

It is easy to see that algebras (\ref{popov_r:mia.defining.equation.1})
(\ref{popov_r:mia.defining.equation.2}) are similar 
to  algebra (\ref{popov_r:mia.olhe}). Therefore, there can exist trasformations 
which, in some sense, reduce (\ref{popov_r:defining.equation.1}) and 
(\ref{popov_r:defining.equation.2}) to (\ref{popov_r:olhe}). 

\medskip

\noindent
{\bf Theorem 4 \cite{popov_r:fshsp}.}
{\it System (\ref{popov_r:defining.equation.1}) is reduced to the system of 
three uncoupled heat equations
$z^a_t=z^a_{xx}, \quad  a=\overline{1,3},$
functions $z^a=z^a(t,x)$ by means of the nonlocal 
transformation
\be \label{popov_r:nonlocal.transformation.1}
g^1=-\frac{z^1_{xx}z^2-z^1z^2_{xx}}{z^1_xz^2-z^1z^2_x}, \quad
g^2=-\frac{z^1_{xx}z^2_x-z^1_xz^2_{xx}}{z^1_xz^2-z^1z^2_x}, \quad
g^3=z^3_{xx}+g^1z^3_x-g^2z^3,
\ee
where $z^1_xz^2-z^1z^2_x\not=0.$}

\medskip

\noindent
{\bf Theorem 5 \cite{popov_r:fshsp}.}
{\it Equation (\ref{popov_r:defining.equation.2}) is reduced
by means of the nonlocal change
\be \label{popov_r:nonlocal.transformation.2}
\theta=-\Phi_t/\Phi_u, \quad \Phi=\Phi(t,x,u)
\ee
and the hodograph transformation
\be \label{popov_r:hodograph.transformation}
y_0=t, \quad y_1=x, \quad y_2=\Phi, \quad \Psi=u,
\ee
to the heat equation for the function $\Psi=\Psi(y_0,y_1,y_2):$
$
\Psi_{y_0}-\Psi_{y_1y_1}=0,
$
where the variable $y_2$ can be asssumed a parameter. }

\medskip

Theorem 4 and 5 show that, in some sense, the problem 
of f\/inding $Q$-conditional symmetry of a PDE is reduced to solving 
this equation, although the def\/ining equations for the coef\/f\/icient 
of a $Q$-conditional symmetry operator are more complicated.

To demonstrate the ef\/f\/iciency of $Q$-conditional symmetry, consider 
a generalization of the heat equation, which is a linear transfer equation:
\be \label{popov_r:linear.transfer.eq}
u_t+\frac{h(t)}{x}+u_{xx}=0.
\ee

\medskip

\noindent
{\bf Theorem 6.  \cite{popov_r:jnmp94,popov_r:umj95}.}
{\it The maximal Lie invariance algebra of (\ref{popov_r:linear.transfer.eq}) 
is the algebra

\vspace{1ex}

1) $A^1=\:\langle u\partial_u, \; f(t,x)\partial_u\rangle \;$ if $\;h\not=\const;$

\vspace{1ex}

2) $A^2=A^1\:+\langle \partial_t, \; D, \; \Pi\rangle \;$ if $\;h=\const,$ 
$h\not\in\{0,-2\};$ 

\vspace{1ex}

3) $A^3=A^2\:+\langle \partial_x+\frac{1}{2}hx^{-1}u\partial_u, \;
G\rangle \;$ 
if $\;h\in\{0,-2\}.$ 

\vspace{1ex}

\noindent Here, 
$f=f(t,x)$ is an arbitrary solution of (\ref{popov_r:linear.transfer.eq}), 
$D=2t\partial_t+x\partial_x,$ 
$\Pi=4t^2\partial_t+4tx\partial_x-(x^2+2(1-h)t)u\partial_u,$
$G=t\partial_x-\frac{1}{2}(x-htx^{-1})u\partial_u.$}

\medskip

Theorems  being like to  Theorems 1--5 were proved for  equation 
(\ref{popov_r:linear.transfer.eq}) too. 

It follows from Theorem 6 that the Lie symmetry of 
equation (\ref{popov_r:linear.transfer.eq}) is trivial in the case 
$h\not=\const.$ But for an arbitrary function $h$,  equation 
(\ref{popov_r:linear.transfer.eq}) is $Q$-conditionally invariant, for example,  
under the folowing operators:
\be \label{popov_r:oper1.2}
X=\partial_t+(h(t)-1)x^{-1}\partial_x, \quad
\tilde G=(2t+A)\partial_x-xu\partial_u, \quad A=\const.
\ee
By means of  operators (\ref{popov_r:oper1.2}), we construct solutions 
of (\ref{popov_r:linear.transfer.eq}):
\[
u=C_2\bigl(x^2-2\int(h(t)-1)dt+C_1, \quad 
u=C_1\exp\left\{-\frac{x^2}{2(2t+A)}+\int\frac{h(t)-1}{2t+A}dt\right\}
\] 
that can be generalized in such a way:
\[
u=\sum_{k=0}^nT^k(t)x^{2k}, \quad
u=\sum_{k=0}^nS^k(t)\left(\frac{x}{2t+A}\right)^{2k}
\exp\left\{-\frac{x^2}{2(2t+A)}+\int\frac{h(t)-1}{2t+A}dt\right\}.
\]
The functions $T^K$ and $S^k$ satisfy systems of ODEs which are easily 
integrated.

Using $Q$-conditional symmetry, the nonlocal equivalence transformation 
in the class of equations of the form (\ref{popov_r:linear.transfer.eq}) 
are constructed too.

In conclusion, we should like to note that the connection between 
$Q$-conditional symmetry and reduction of PDEs is analogous 
to one between Lie symmetry and integrating ODEs of the f\/irst order. 
The problem of f\/inding $Q$-conditional symmetry is more complicated 
than the problem of solving the initial equation. But constructing 
a $Q$-conditional symmetry operator by means of any additional 
suppositions, we obtain reduction of the initial equation. 
And it is possible  that new classes of symmetries, which 
are as many as $Q$-conditional symmetry and will be found 
as simply as the Lie symmetry will be investigated in the future.

\medskip

{\sl The author is grateful to Prof.~R.Z.~Zhdanov and 
Prof.~I.M.~Tsyfra for valuable discussions.}

\medskip

This work is supported by the  DFFD of Ukraine (project  1.4/356).

\label{popov_r-lp}

\end{document}